\documentclass[journal]{vgtc}                % final (journal style)
\usepackage{mathptmx}
\usepackage{graphicx}
\usepackage{times}
\let\ifpdf\relax
\usepackage{epstopdf}
\usepackage{afterpage}% http://ctan.org/pkg/afterpage
\usepackage[bookmarks,backref=true,linkcolor=black]{hyperref} %,colorlinks
\hypersetup{
  pdfauthor = {},
  pdftitle = {},
  pdfsubject = {},
  pdfkeywords = {},
  colorlinks=true,
  linkcolor= black,
  citecolor= black,
  pageanchor=true,
  urlcolor = black,
  plainpages = false,
  linktocpage
}
\newcommand{\hidecomment}[1]{}
\onlineid{0}

\vgtccategory{Research}

\vgtcinsertpkg

\title{RioBusData: Outlier Detection in Bus Routes of Rio de Janeiro}

\author{Aline Bessa$^{1}$, Fernando de Mesentier Silva$^{1}$, Rodrigo Frassetto Nogueira$^{1}$, Enrico Bertini$^{1}$, and Juliana Freire$^{1,2}$ % <-this % stops a space
\thanks{The first three authors have contributed to the work equally.} % <-this % stops a space
\thanks{$^{1}$Department of Computer Science and Engineering, New York University} 
\thanks{$^{2}$New York University Center for Urban Science \& Progress} }

\abstract{Buses are the primary means of public transportation in the city of Rio de Janeiro, carrying around 100 million passengers every month.~\footnote{Source in Portuguese: \url{http://www.rioonibusinforma.com/site/wp-content/uploads/2015/01/TB4-Dados-operacionais-mensais-do-municipio-do-rio-de-janeiro-ano-de-2014.pdf}.} Recently, real-time GPS coordinates of all operating public buses has been made publicly available -- roughly 1 million GPS entries each captured each day. In an initial study, we observed that a substantial number of buses follow trajectories that do not follow the expected behavior. In this paper,  we present RioBusData, a tool that helps users identify and explore, through different visualizations, the behavior of outlier trajectories. We describe how the system automatically detects these outliers using a Convolutional Neural Network (CNN) and we also discuss a series of case studies which show how RioBusData helps users better understand not only the flow and service of outlier buses but also the bus system as a whole.}


\begin{document}

\firstsection{Introduction}

\maketitle

Recently, the city of  Rio de Janeiro released the real-time GPS coordinates of all their operating public buses alongside their timestamps~\footnote{Data available at 
http://data.rio.rj.gov.br/dataset/pontos-dos-percursos-de-onibus}. 
These data allow a broad range of monitoring and analysis applications for Rio's public bus system.
In an initial study, we observed that many buses behave anomalously: their trajectories do not follow the expected (planned) route.
Given the scale and complexity of the data, there are many challenges involved in both identifying and understanding these anomalies.
While there have been approaches to visualize trajectories, manually identifying anomalous trajectories in such a large collection is not practical -- there are almost 1 million GPS entries each day.
In this paper, we propose the use of Convolutional Neural Networks (CNN)~\cite{lecun1998gradient} to automatically detect outliers. We consider as an outlier any trajectory that contains at least one GPS entry that does not follow the expected behavior in space (e.g., buses outside their planned routes) and time (e.g., delayed buses). 
We have experimented with traditional event-detection techniques, including K-means and K-nearest neighbors, and we show that CNNs outperform these techniques both in performance and detection accuracy. 
CNNs are one of the most popular deep-learning methods. A CNN models high-level abstractions in data by using model architectures with complex structures or comprised by multiple non-linear transformations~\cite{lecun1998gradient}.  By using CNNs, it is possible to derive outlier models for large volumes of data, a task that is time consuming and not feasible to be accomplished manually. 

As we discuss in Section~\ref{label:related_work}, while the exploration spatio-temporal data sets has received substantial attention in the visualization community, the same cannot be said of outlier detection. In spatio-temporal data sets, it is particularly challenging to analyze these outliers. By inspecting the values of the GPS coordinates, it is hard to know where they are in the map, or whether they lie in a proper street.  To help users explore and understand these outliers, we have designed RioBusData, a visual analytics system that combines automatic outlier detection with insightful visualizations.

\hidecomment{
In the context of this work, an outlier bus is any bus with at least one GPS entry identified as an outlier -- i.e., not following its expected behavior -- by a Convolutional Neural Network 
(CNN)~\cite{lecun1998gradient} especially designed for outlier detection. Outlier buses can be either spatial --  e.g., buses running outside their average route (far from where they 
were supposed to be) -- or temporal -- e.g., delayed buses.  There are different ways to detect outliers. One can argue that just by visualizing the buses (on a map, for example), 
it is possible to recognize which of them are outliers. From our experience, it is not so simple to proceed in this way: almost 1 
million GPS entries per day are generated by Rio's bus system, and a reasonable amount of them are outliers. To detect outliers in a large-scale fashion just by looking at them is 
thus very time-consuming. Finally, and most importantly, if the users have to identify the outliers manually, 
it takes more time for them to analyse and understand what is going on with the buses. 
These observations led to the need of having a model to detect outliers automatically. We analyzed different models for outlier detection and  CNNs were the best studied option both in terms of 
performance and detection accuracy. 
}

\hidecomment{The main contribution of our work is to provide a tool that, through the application of visual analytics techniques, enables users to identify and explore outliers detected in a spatio-temporal data set. }

The remainder of the paper is organized as follows. 
We discuss related work on Section~\ref{label:related_work}. Section~\ref{label:dataset_tasks} provides a description of the bus trajectory data set and a discussion of the tasks and analytical questions needed for outlier exploration in this setting. Our approach to design a CNN for outlier detection is presented in Section~\ref{section:cnn}. The visualizations and user interface for RioBusData are described in  Section~\ref{label:visualization_and_interaction_design}.
In Section~\ref{label:findings_and_insights}, we discuss use cases for RioBusData alongside findings and insights obtained by the system. We conclude in
Section~\ref{label:conclusions_and_future_work}, where we outline directions for future work. 

For an online demo for RioBusData, see \url{http://rodrigonogueira4.github.io/BusData/Outlier_Vis/index.html}.

\section{Related Work}
\label{label:related_work}

\hidecomment{In this section we present some works that are related to RioBusData. Some works tackle the problem of trajectories visualization; some address visualization systems for outlier detection
 and analysis; and other papers are associated with Deep Learning applications for outlier detection. }

\textbf{Visualization of Trajectories} The visualization of trajectories is a well explored problem inside the visualization community.  Wegenkittl et al., for example, proposed different methods, based on parallel coordinates, to tackle the visualization of trajectories in high-dimensional dynamical systems~\cite{wegenkittl1997visualizing}. Although the resulting tool was used in many different applications, it is not suitable for visualizing a very large number of samples as required for the bus data.  
Tominski et al. also proposed a tool for the visualization of trajectories~\cite{tominski2012stacking}. Through a hybrid 2D/3D display, trajectory bands are stacked on top of a map that is shown for space context. This approach provides a solution to the simultaneous exploration of both temporal and spatial attributes. However, it is not clear how this approach can be adapted to our problem, where the analysis of outlier behaviors is the main focus. Ferreira et al. proposed TaxiVis, a system that supports interactive querying and exploration of spatio-temporal data through visualization~\cite{ferreira2013visual}.  They showed the effectiveness of their system for taxi data consisting of spatial, temporal, and other attributes (e.g., fare, trip length) associated with taxi trips. Although TaxiVis does not handle  outlier detection or exploration, some of its features served as inspiration for the design of RioBusData.
Barbosa et al. proposed Vistradas, a visual analytics system that supports the exploration of bus trajectory data~\cite{barbosavistradas}. Vistradas allows users to analyze bus uniformity, verify  bus routes and the impact of events in bus traffic. RioBusData and Vistradas are similar in that they use the same data set and support similar tasks. However, the focus of RioBusData is on outlier detection, an issue not addressed by Barbosa et al. 

\textbf{Visualization Systems and Tools for Outlier Detection} While the visualization of trajectories has been extensively studied by the visualization community, to the best of our knowledge outlier detection and visualization have received comparatively much less attention.

Novotny and Hauser~\cite{novotny2006outlier} presented a method for focus+context visualization in parallel coordinates where, through the use of a binning and filtering algorithm, certain points are detected as outliers. This work differs from RioBusData in significant ways. First, the outlier detection is performed through the visualization. In contrast, RioBusData allows users to visually explore previously identified outliers. Second, detected outliers are not correlated, whereas in RioBusData we deal with trajectories consisting of a sequence of (related) GPS entries. Last, but not least, RioBusData was designed for the exploration of spatio-temporal data sets.

More closely related to RioBusData is the work by Sekhar et al.~\cite{sekhar2002data}. They  described a web-based visualization package that summarizes spatial patterns and temporal trends.  They proposed data mining algorithms for filtering out data sets to identify spatial outlier patterns which, like RioBusData, were implemented and tested using a real-world
traffic data set. RioBusData is different from this work in the sense that the outliers are previously and automatically detected with a Machine Learning model, and the visualizations are used to inspect, understand, and take actions based on the already processed information. Finally, Sun et al. propose a simulation-based method that helps in the visual detection of outliers in spatio-temporal data by adjusting functional boxplots~\cite{sun2012adjusted}. This work is also different from RioBusData in the sense that the detection of outliers is one of the goals of the visualizations. Besides, the proposed method is not suited for large, continuous streams of data, which is the case of the data set explored in our work.

\textbf{Deep Learning for Outlier Detection} The detection of outliers in temporal series has not been widely explored in the context of Deep Learning algorithms. However, various works already demonstrated the potential of neural network techniques for the time series tasks. In~\cite{dalto2015deep}, for example, deep neural networks were applied for ultra-short-term wind prediction, and the results showed that deep neural networks outperform shallow architectures.

The deep models are compared to a classical Dynamic Time Warping (DTW) approach and the results indicated that the deep model is not only more efficient (especially for large data sets) than the state of the art, but also yielded better accuracy for two standard benchmarks.
More closely related to our approach is the work by Hawkins et al.~\cite{hawkins2002outlier}. They used Replicator Neural Networks (RNNs) to reconstruct the input sample and, once the model was trained, samples that have a high reconstruction error were marked as outliers. Our approach to outlier detection is similar to this work in the sense that both models learn only frequent samples, while uncommon samples yield higher errors. On the other hand, RioBusData does not use a sample reconstruction approach: instead of applying RNNs over the data, it uses CNNs. To the best of our knowledge, RioBusData is the first application of CNNs for outlier detection, and, as outlined in Section~\ref{section:cnn}, the results are promising.

\section{Domain Characterization}
\label{label:dataset_tasks}

The design of RioBusData was motivated by the need to explore bus route data from Rio de Janeiro. In what follows, we describe the Rio bus data set and present the task abstractions supported by the system. We should note that the system is flexible and can be applied to other data sets and trajectory data. 

\hidecomment{has a versatile design and would potentially work for spatio-temporal data sets related to other cities or even for vehicles other than buses. Nevertheless, at this moment it is an application restricted to Rio de Janeiro's citizens because this was the data we had available for our studies at the time. In this section, we present a characterizations of this data set and then we introduce the task abstractions addressed by RioBusData.  }

\subsection{Data Set Description}

The \emph{bus data set} used in this paper consists of GPS entries (records) for over 9,000 buses in transit in Rio de Janeiro, covering approximately 490 bus lines, from September 26, 2013 to January 9, 2014. It contains 151,730,254 records, totaling approximately 23GB. Each record contains: the bus ID, the bus route ID, timestamp in UTC format, latitude, and longitude. We added two attributes to these records: (i) the total trip time -- from the beginning to the end of the line route; and (ii) whether or not it was identified as an outlier by the CNN. In this section we explain how trip times are computed. Details of how our CNN works and detects outliers are given in Section~\ref{section:cnn}.

The main purpose of computing bus trip times is to help users better understand temporal outliers. If a GPS entry seems to be associated with an outlier bus, the user can check its trip times and analyze whether it is moving too slow or too fast, when compared with other buses. To compute the total time it takes for one bus to go from the beginning to the end of its route, we had to first identify their initial and final bus stops. In order to do so, we used an auxiliary GTFS data set that contains the bus stops for all the 486 bus lines in Rio de Janeiro. Each entry in the GTFS data set has a bus route ID corresponding to a bus line, a sequence number indicating what bus stop the entry corresponds to, and its latitude and longitude. With this information we compute, for each entry in the bus data set, how far it is from its initial bus stop.
Given a sequence of consecutive entries in the bus data set, with fixed bus ID and bus route ID, we can sort them by their timestamps and compute how far they are from their initial bus stop. Every time we find a cycle -- i.e., a set of points getting farther from the initial bus stop until they start to get closer -- we identify it as a trip from the beginning to the end of the route and compute its associated time by using the timestamps of the entries in the cycle. This is a simple heuristic to estimate bus trip times, but our interactions with RioBusData interface indicate that it works quite well in practice.

\subsection{Task Abstractions}
\label{subsec:task_abstractions}

While analyzing outliers, users need to be able to both distinguish and understand the outliers. 
\hidecomment{
There are a couple of task abstractions for the problem of detecting outliers in bus trajectories. Some of them are related to the analysis of outliers in general; others have to do with the differentiation and understanding of spatial and temporal outliers. In this subsection we describe these task abstractions, and in Section~\ref{label:findings_and_insights} we describe use cases where they are addressed by RioBusData.}
RioBusData supports a set of task abstractions that help users answer the following questions:
\hidecomment{
We pose our task abstractions as a set of analytical questions.
As for \textbf{outliers in general}, RioBusData helps users come up with answers to the following questions: }

\begin{itemize}
\item How many outliers are there in the bus data set?
\item Which days and hours have more outliers?
\item Which bus lines have more outliers? 
\end{itemize}

With respect to \textbf{spatial outliers}, RioBusData addresses the following 
questions: 

\begin{itemize}
\item What bus lines have more spatial outliers?
\item How do these outliers stray away from their routes, and where?
\item What buses are so off their routes that they are probably incorrectly labeled in the data set? 
\end{itemize}

Finally, with respect to \textbf{temporal outliers}, RioBusData provides different mechanisms for users to explore the following questions:

\begin{itemize}
\item What bus lines have more temporal outliers?
\item Are these temporal outliers usually running faster or slower than their expected average speed?
\item In what dates, or hours, do these lines get more affected by temporal outliers?
\end{itemize} 

\section{Detecting Outlier Buses with a Convolutional Neural Network}
\label{section:cnn}

In this section we explain how we designed the CNN used for the bus data and how it detects
outliers. Before selecting CNNs for outlier detection, we tried using simpler, standard approaches, namely K-means and K-nearest neighbors. K-means is a popular method for cluster analysis in Data Mining~\cite{data-mining}, and used it to cluster trajectory data for all different bus route IDs present in the bus data set. Those GPS entries that were not inside the proper cluster, or were too far from their proper centroids, were considered outliers. This method did not scale well for our problem and the outliers it detected did not make much sense.

The results could  potentially be improved by tuning different parameters, but given that the computations were too slow, we decided against using K-means. We then tried using K-nearest neighbors, a fast classification method commonly used in pattern recognition applications~\cite{data-mining}. Since most GPS entries in the bus data set are near bus stops, we decided to use the GPS coordinates of bus stops present in the GTFS data set, as references for how bus trajectories should be, and then we let a K-nearest neighbors algorithm detect which entries do not follow these patterns. The results were interesting for spatial outliers, but adapting the algorithm to detect other inadequate behaviors would not be trivial. 

To detect temporal outliers using K-nearest neighbors, we would have to modify a significant part of its implementation. If we found  other interesting types of outliers, additional modifications would be required.  Ideally, a method for outlier detection should be flexible  and able to support different kinds of outliers.

We then decided to try Convolutional Neural Networks, 
given its recent success in several different Data Mining applications and its ability to learn non-linear patterns in a data set~\cite{lecun1998gradient}. To the best of our knowledge, 
RioBusData is the first application of CNNs for outlier detection.

A classical CNN is composed of alternating layers of convolution and local pooling (i.e., subsampling)~\cite{lecun1998gradient}. The aim of the convolutional layer is to extract patterns found within local regions of the input sequence, by convolving a template over it, and outputting this as a feature map $c$, for each filter in the layer. A non-linear function $f(c)$ is then applied element-wise to each feature map.
The resulting activations $f(c)$ are then passed to the pooling layer. This layer then aggregates the information within a set of small local regions, $R$, producing a pooled feature map $s$ (normally of smaller size) as the output. 
Among the various types of pooling, max-pooling is the most commonly used, and it selects the maximum value of each local region $R$.

\begin{figure}
\centering
\includegraphics[width=0.25\textwidth]{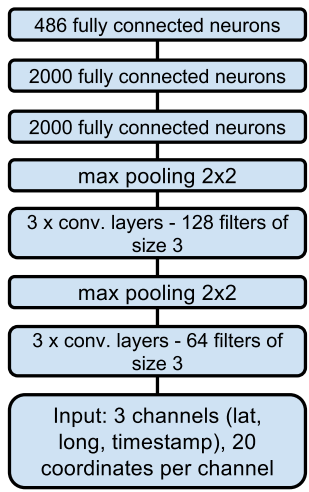}
%\vspace{-2cm}
\caption{\label{fig:cnn_architecture}Illustration of the CNN architecture used in this work. The input to the network are 20 GPS coordinates of single bus that are coded in three channels (1-D vectors), which corresponds to the timestamp, latitude and logitude of each GPS coordinate. This input is fed to the three convolutional layers, one max pooling layer, three convolutional layers, one max pooling layer and then three fully connected layers. The last layer predicts the line (among 486 possibilities) that the input 20 GPS coordinates belongs to.}
\end{figure}

For RioBusData, we started off by picking a standard CNN implementation provided by Caffe, a Deep Learning framework~\footnote{Caffe's implementations and documentation are available at 
\url{http://caffe.berkeleyvision.org/}}. We then modified it to better suit our needs, and the resulting network 
 is illustrated in Figure~\ref{fig:cnn_architecture}. It is composed of six convolutional layers with 1-D filters of size 3, two max-pooling layers with stride of 2  
and two fully connected layers with 2000 neurons each. 
The last layer is a fully connected layer with 486 neurons, and it corresponds to the number of classes (in this case, bus route IDs) 
that the network will predict.

To compute whether or not a GPS entry corresponds to an outlier, our CNN is trained over 90\% of the bus data set.  In the training phase, the CNN receives as input a sequence of 20 consecutive GPS entries for a particular bus ID and bus route ID. That usually corresponds to a trajectory of approximately 2 hours.  Each GPS entry in the sequence is reduced to 3 channels -- its longitude, latitude and timestamp, and each channel is normalized to have mean equals to zero.  The CNN is trained in a supervised fashion and predicts the bus route ID that each sequence of points belongs to. Thus, given that there are 486 distinct route IDs (lines) in the Rio de Janeiro bus system, the network outputs the predicted probability for 486 classes. The architecture of the network is illustrated in Figure~\ref{fig:cnn_architecture}.

We validated models with different learning rates over the remaining 10\% of the bus data set. 
The validation was carried out in order to maximize route ID prediction accuracy.
It is important to mention that a few GPS entries are not associated with correct route IDs in the training and validation sets. These entries are infrequent and are learned by the CNN as outliers.  Highly-frequent patterns in both sets are learned as parts of standard routes and GPS entries presenting such patterns are unlikely to be learned as outliers.  Currently, our best model has an accuracy of 94.9\%.

After the training and validation phases finish, the outliers are predicted by passing sequences of 20 consecutive GPS entries, with fixed bus ID and bus route ID, to the trained model. The model then outputs a score 
ranging from 0 to 1 for each one of the 486 possible bus route IDs. The bus route ID associated to the highest score is the line that the entries in the sequence belong to. The GPS entries in a sequence are considered outliers if the largest score is below a certain threshold -- 0.1 in our preliminary experiments -- or the predicted bus route ID is different from the one assigned to them in the bus data set. We use sequences of GPS entries instead of each entry individually because it would be considerably harder to detect outliers, especially inadequate temporal behaviors, by using each entry separately. In a sense, certain outlier behaviors cannot be easily noticed when one does not analyse sequences of points at the same time.

\section{Visualization and Interaction Design}
\label{label:visualization_and_interaction_design}

\begin{figure*}[!ht]
\centering
\includegraphics[width=\textwidth]{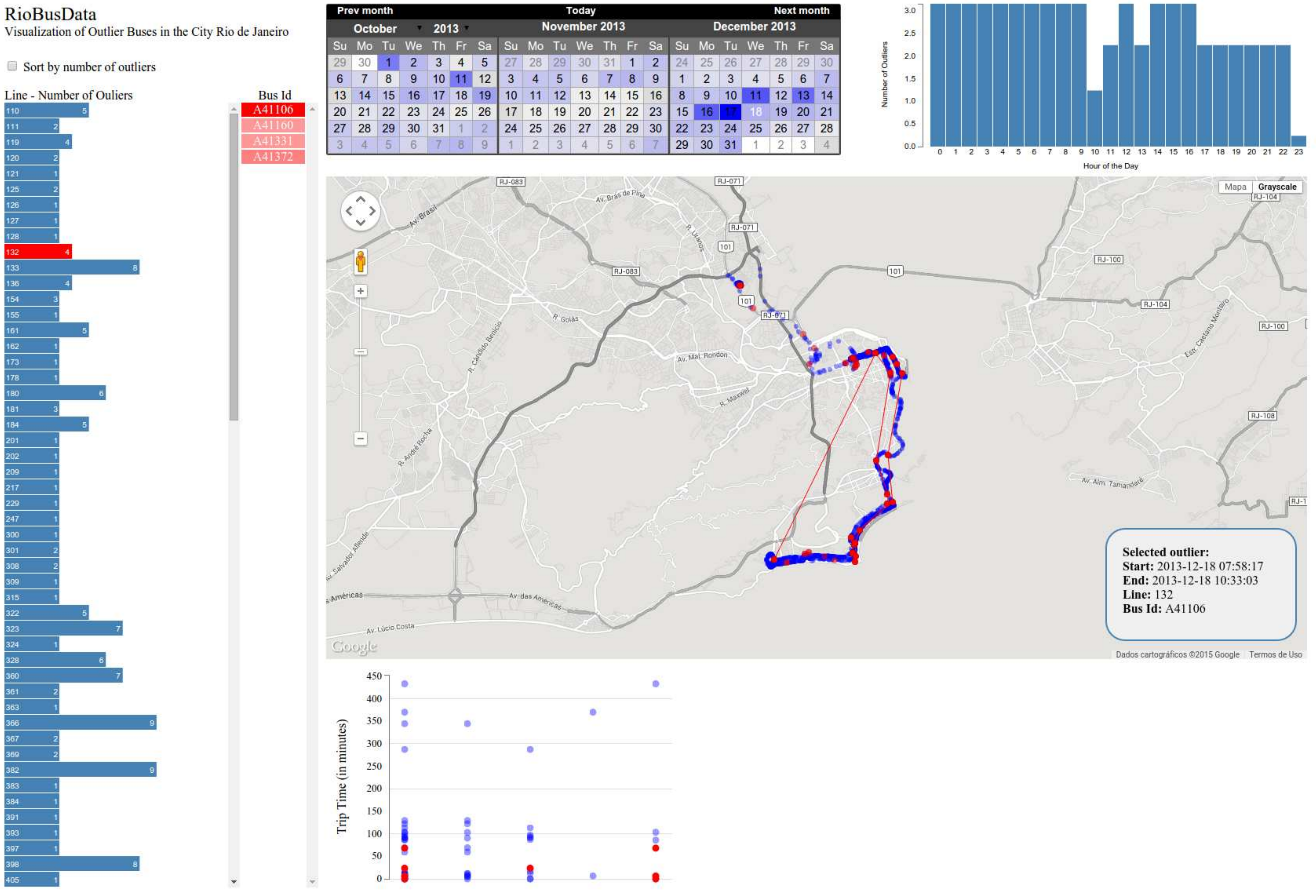}
\caption{\label{fig:overview}An overview of RioBusData. The leftmost view (View 1) corresponds to the bus line chart, with a bar for each bus line and its corresponding 
number of detected outliers. The calendar on the top (View 2) indicates the number of detected outliers per day: the darker a day is, the more 
outliers it has. The hour chart on the right (View 3) displays the number of detected outliers for different hours -- 0 corresponds to 12AM and 23 corresponds to 11PM. The map 
in the middle (View 4) is where GPS entries are plotted: the red ones correspond to outlier buses while the blue ones are associated with regular buses. A  filter per bus ID 
to the left of the map allows users to analyse one bus at a time. Finally, the trip time chart on the bottom (View 5) allows users to check trip times for different bus IDs and investigate 
whether they are associated with outlier buses.}
\end{figure*}

RioBusData provides a set of views that allows users explore outliers in the bus data. An overview of the user interface is shown in Figure~\ref{fig:overview}.

The design of RioBusData follows the Shneiderman's mantra: overview first, zoom and filter, then details-on-demand. In this way, users can deal with the large volume of information in a principled fashion. When users start interacting with RioBusData, they see all the detected outliers highlighted in the bus line chart in View 1, the calendar in View 2, the hour chart in View 3, and the map in View 4. This overview already provides important information about the bus data set, e.g., which days and hours have more outliers, where most outliers are geographically, and which lines are associated to more outliers. As we explain later in this section, the trip time chart in View 5 gets populated after the user interacts with the other views.

These five views proved to be sufficient for a simple yet effective analysis of outliers in the public bus transportation system. Views 1, 2, and 3 inform the users about the distribution of outliers by line, date and hour respectively. They also work as filters for the data, as users can choose to analyse specific bus lines, ranges of days or hours by interacting with them. Outliers and regular buses plotted in View 4 will always respect the filters set by previous interactions with Views 1 to 3. Users can then ask for details about a particular outlier plotted in View 4. The interaction with an outlier in View 4 populates View 5, where its trip time can be inspected. The next subsections detail each view of RioBusData and discuss some of its main design decisions.

\subsection{Bus Line Chart (View 1)}

To help users answer questions such as \emph{Which bus lines have more outliers?}, we had to aggregate all outliers per line and display it. Initially, we implemented a 
scatterplot where each point represented a different bus line, axis X corresponded to the number of GPS entries per line, and axis Y to the number of detected outliers per line. The 
resulting scatterplot was crowded, requiring zooms to enable differentiation of points that were too close to each other. To accomodate a clearer view of all bus lines, the scatterplot would have 
to fill most of the space on the screen, imposing a visualization where not all views would be accessible at the same time. 
Thus, instead of a scatterplot, we used a bar chart representation -- the bus line chart.
The bus line chart informs the distribution of outliers in the bus data set and also works as a filter. It can be used to select a specific bus line whose GPS entries will be plotted on the map in View 4.  The bus line chart encodes bus line IDs as the leftmost number in each bar and the respective number of detected outliers as the rightmost number. Its relevant visual mark is the size of each bar, which corresponds to the number of outliers in the associated bus line. It is also important to mention that there is a checkbox on top of the bus line chart with which the bars can be sorted according to the number of outliers.

\subsection{Calendar and Hour Chart (Views 2 and 3)}

According to the task abstractions in Section~\ref{subsec:task_abstractions}, users should be able to gather hypotheses for questions as \emph{Which days have more outliers?}. Initially, we addressed this by displaying a histogram with each bar corresponding to one day and its height corresponding to the number of detected outliers. Considering that we have approximately 90 days of data available, the resulting view was unnecessarily large. The users had to scroll through the histogram to access its different parts. Consequently, we moved to an actual calendar, with different hues of blue as a visual mark indicating the number of outliers for each day -- the darker the hue of blue the more outliers in the associated day. Besides being more compact than a histogram, the calendar at View 2 allows comparisons between different days of the week in a more straightforward fashion due to their natural week alignment.
 
The calendar in View 2 also works as a filter for the bus data. By clicking on a day, or on multiple  days, the GPS entries plotted in View 4 will be associated to the selected days exclusively.

In RioBusData, it is also possible to answer questions such as \emph{Which hours have more outliers?}. To do so, users just have to interpret the information contained in the hour chart in View 3. Initially, we had two sliding controls to filter GPS points by start time and end time, in a range of 24 hours. These controls give users a finer grained filtering, but are less intuitive to manipulate than the hour chart. In the hour chart, users just have to click on one or more bars to filter GPS entries by specific hours. Furthermore, just by observing the height of different bars, the main visual mark in this chart, users can understand which hours are associated to more outlier buses. Before, with the sliding controls, users had to discriminate different time ranges and observe the amount of red points plotted on View 4 to estimate which hours had more outliers. The hour chart is thus much more straightforward for comparisons among different hours.

\subsection{Map (View 4)}

The map view is the core of RioBusData, where the most relevant piece of information is represented: regular buses as blue points and detected outliers as red points. The points plotted in View 4 may be filtered by bus line, different days or hours. If no filter is used, a uniformly subsampled set of all outliers detected in the bus data set is plotted.  Buses plotted on the map can be further filtered by the bus IDs, which are outlined in a small chart to the left of the map, in View 4.

Outliers in the visualization that are far from blue points are spatial outliers. They fail to  follow their planned routes.
Buses in Rio are required to travel in specific routes and to provide their service in a timely manner. Unfortunately, buses may alter their services due to a number of reasons such as roadblocks or driver misconduct (it is not uncommon for bus drivers of the same line to race each other during their shifts). Also, GPS data capture is active when buses are moving to their garage and when they are stopped, and in these occasions these buses are also behaving as spatial outliers. 
When outliers are mixed with blue points they are usually temporal outliers. In other words, they are following their routes correctly, but there is something wrong with their speeds.  Buses can behave as temporal outliers for several reasons, including traffic jams, break downs, mistakes made by the bus drivers. It may also be the case that a specific bus is emitting GPS entries with incorrect timestamps, thus giving the impression that it is later or earlier than it actually is. Finally, it is important to note that there is a correlation between spatial and temporal outliers, as buses which stray from their routes will usually take more time to hit their next stops. It is very straightforward to verify spatial outliers in the map, but to better analyse temporal ones, a different visual representation is necessary.

\subsection{Trip Time Chart (View 5)}

By following the Shneiderman's mantra, RioBusData allows users to ask for details about specific outliers by clicking on their point representations on the map in View 4. When an outlier point is selected, the trip times for buses of its line are showed in the trip time chart in View 5. Trips that have at least one outlier point are displayed in red and other trips in blue. The Y-axis corresponds to  the number of minutes it took for a trip to complete and the X-axis separates trips by showing a plot of all trips made by buses of that line, with the label \emph{All buses}, and plots for individual buses of that line that have at least one trip with outlier points. Before implementing the trip time chart, we used a line chart to compare the time of different buses of a same line running on a segment of their route where an outlier had been found. Not only was the graph too crowded, but the data itself was not suitable for this type of analysis, rendering it ineffective.  Consequently, to better understand whether an outlier was of spatial or temporal nature, we computed the trip times of buses and through the trip time chart it is possible to compare the time of all buses on the same line and guide an assumption of the relationship between outliers and the time they take to complete trips (which can either be much less or much more than average).

\section{Use Cases}
\label{label:findings_and_insights}

In this section we show, through different use cases, how RioBusData addresses the analytical questions presented in Section~\ref{subsec:task_abstractions}. We also describe findings and insights obtained by interacting with the different components of the system.

\textbf{Use Case 1: Getting to know the outlier buses.} Our first use case involves users who want to understand how outlier buses are distributed. The components explored in this use case are illustrated in Figure~\ref{fig:overview} at 
Views 1, 2 and 3, and the related analytical questions are:

\begin{itemize}
\item How many outliers are there in the bus data set?
\item Which days and hours have more outliers?
\item Which bus lines have more outliers?
\end{itemize}

In the bus line chart, by selecting the 
\emph{Sort by number of outliers} box, users can quickly verify which bus lines have the larges number of outliers. Although it is expected that outliers occur in lines that serve the city 
center (where more atypical events are likely to occur), a large  number of outliers might be associated with poor road infrastructure or an inadequate bus service -- an insufficient number of buses to serve the population. 

The systems allows users to compare different days, in terms of numbers of outliers, by looking at the day calendar. December is the month with most outliers (see e.g., Dec 17th) while November has some days with no detected outliers (e.g., Nov 20th). One of the hypothesis behind this is that December coincides with the beginning of Summer vacation in Brazil, which is correlated with a big influx of tourists and a resulting increase in traffic. 
Another potential explanation could be the fact that December is the holiday season. It is important to point out that the number of GPS entries does not vary during the three months of data we studied, while the number of outliers does. This indicates that that more outliers are not necessarily correlated with more GPS entries, probably reflecting actual 
disruptions in traffic. 
Furthermore, according to the hour chart, most outliers lie between 12AM and 1AM. Users would probably expect more outliers during rush hour, but it turns out that, between 12AM and 1AM, several buses are going back to their garages, generating a high number of spatial and temporal outliers. In the particular case of routes involving garages, we decided to consider them as outliers because they do not follow the standard behavior learned for their corresponding routes. Nonetheless, it is important to mention that the CNN might have also learned to distinguish between regular routes and routes to and from the garage. This requires further analysis.

\textbf{Use Case 2: Exploring Temporal Outliers}

Users willing to study temporal outliers can choose different bus lines in the bus line chart and observe which of them corresponds to more outliers intertwined with regular buses in the map of View 4. When outliers are mixed with blue points, they are not far from their regular routes and consequently are likely to be temporal outliers. By clicking on a temporal outlier on the map, its corresponding trip times are plotted in the trip time chart, allowing users to investigate why they were classified as temporal outliers. If trip times are considerably longer than the average for regular buses, it may be moving too slowly. By interacting with the calendar and the hour chart, and analyzing different numbers of temporal outliers on the map, users can also study which days and hours are more likely to get affected by them. These interactions help users tackle the following analytical questions:

\begin{itemize}
\item What bus lines have more temporal outliers?
\item Are these temporal outliers usually running faster or slower than their expected average speed?
\item In what dates, or hours, do these lines get more affected by temporal outliers?
\end{itemize} 

\begin{figure}
\centering
\includegraphics[width=0.23\textwidth]{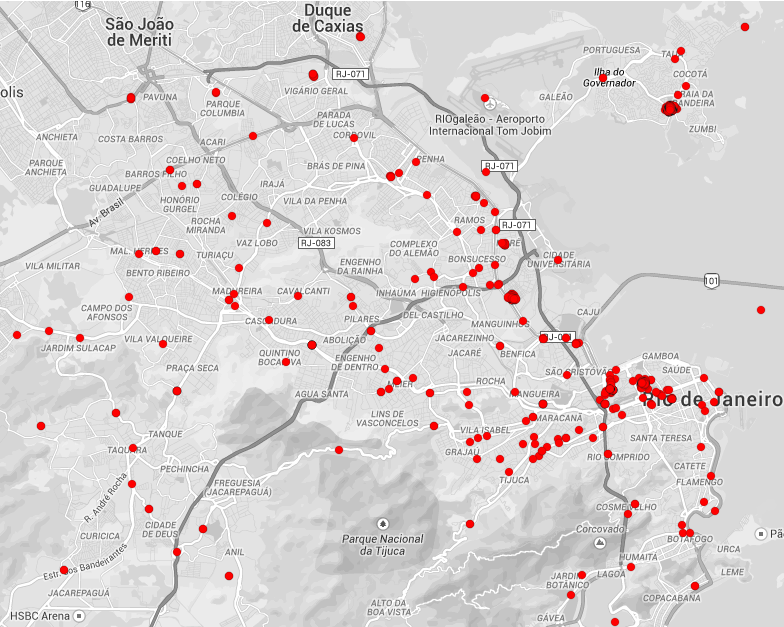}
\includegraphics[width=0.23\textwidth]{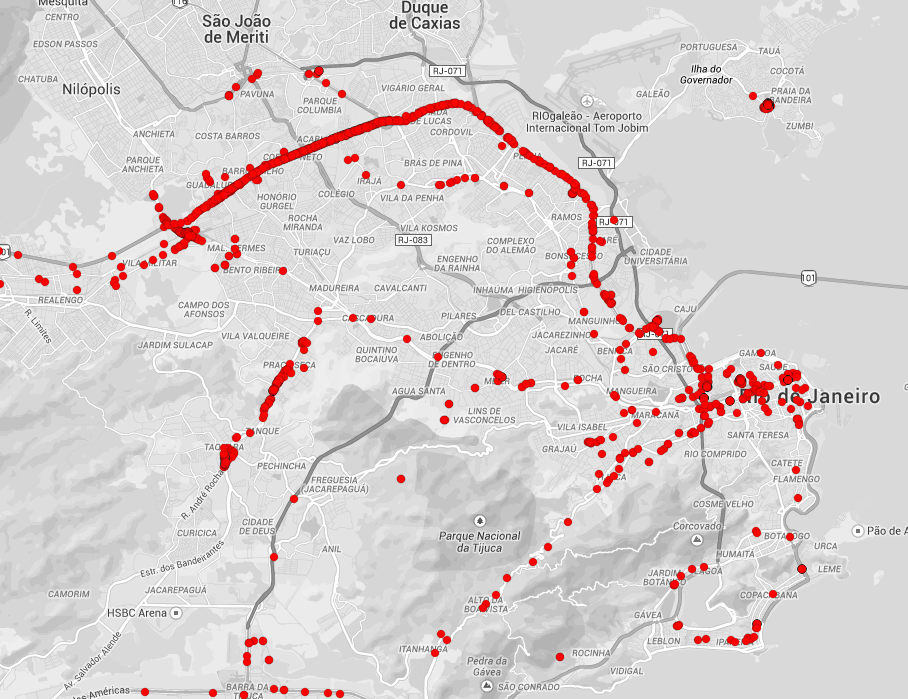}
\caption{\label{fig:trafficjam}Left: Outliers from all lines plotted on a regular day. Right: Atypical day when multiple outliers were detected on some of the main avenues of Rio de Janeiro, 
indicating traffic jams.}
\end{figure}

A common pattern of temporal outliers consists of traffic jams, as showed in Figure~\ref{fig:trafficjam}. This figure shows all outliers for a regular day on the left,
 and outliers for a day with congested roads on the right. In the latter, outliers follow the regular route determined for the bus line but which, due to the volume of traffic, 
are running in a very slow pace.
This shows that one of the main avenues of Rio de Janeiro contains a higher density of outliers than that of surrounding areas.  This example also highlights that RioBusData can be used to detect traffic jams from indirect measurements: number of outliers.

\begin{figure}
\centering
\includegraphics[width=0.44\textwidth]{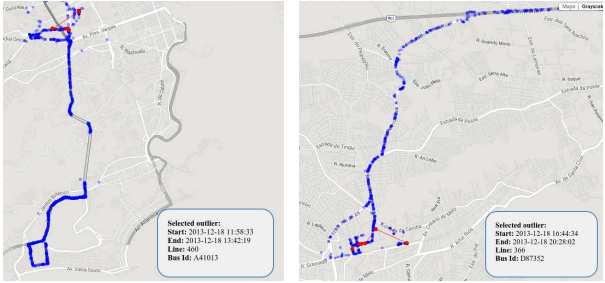}
\caption{\label{fig:brokenbuses}Two examples of buses that stood for many hours in some streets of Rio de Janeiro. This is a clear example of temporal outliers.}
\end{figure}

By interacting with RioBusData, it is also possible to identify temporal outliers that correspond to broken buses. For instance, Figure~\ref{fig:brokenbuses} shows outlier buses that were stuck for more than one hour in busy streets of the city. There are at least two hypotheses behind such pattern: either the GPS sensors of such buses are broken, and they may actually be running properly, or  they actually broke while in service or stopped for some other reason.

\textbf{Use Case 3: Exploring Spatial Outliers} 

Spatial outliers are more straightforward to identify, as their corresponding points on the map (View 4) are physically far from the buses with normal behavior. However, RioBusData provides mechanisms for more interesting, non-trivial analyses involving spatial outliers. By selecting different lines in the bus line chart and examining the map, for example, it is possible to get a an idea of which lines are more affected by spatial outliers. Similar comparisons can be performed by interacting with the calendar and the hour chart. Furthermore, if certain outlier buses seem to be following another bus route, users can hypothesize whether or not their actual bus route IDs are correct in the bus data set. Through these interactions, users can devise answers and hypotheses for the following analytical questions:

\begin{itemize}
\item What bus lines have more spatial outliers?
\item How do these outliers stray away from their routes, and where?
\item What buses are so off their routes that they are probably incorrectly labeled in the data set? 
\end{itemize}

\begin{figure}
\centering
\includegraphics[width=0.32\textwidth]{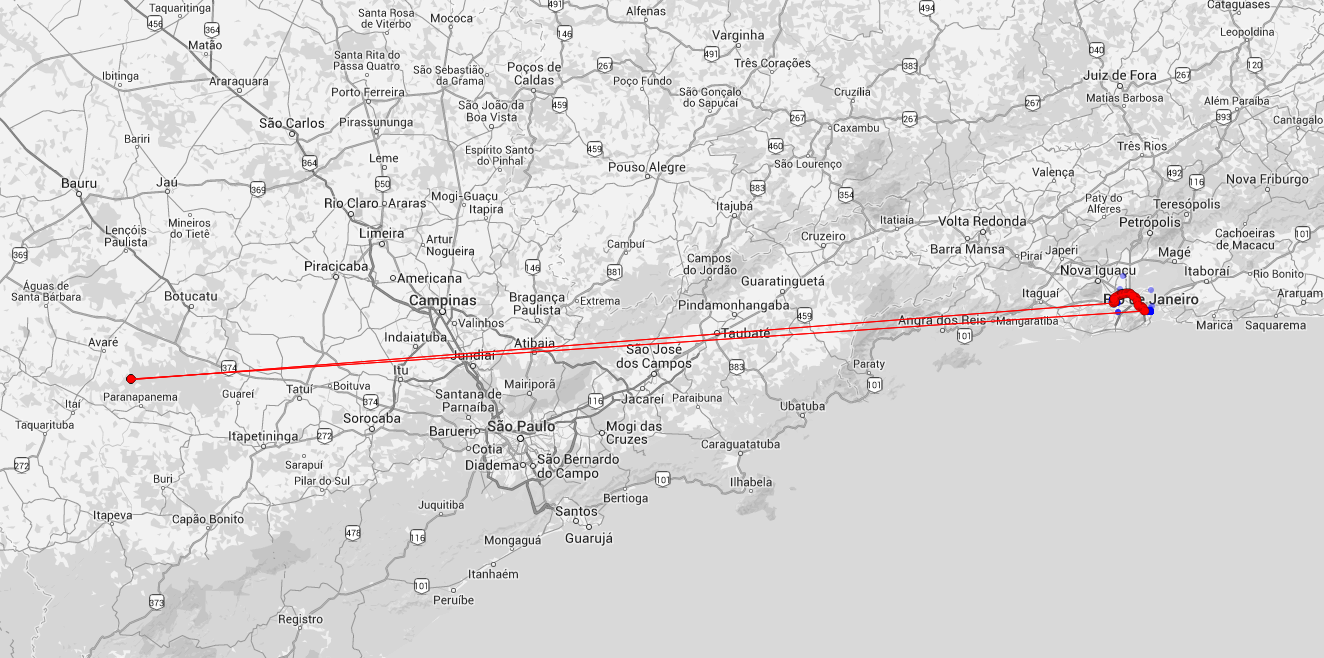}
\includegraphics[width=0.14\textwidth]{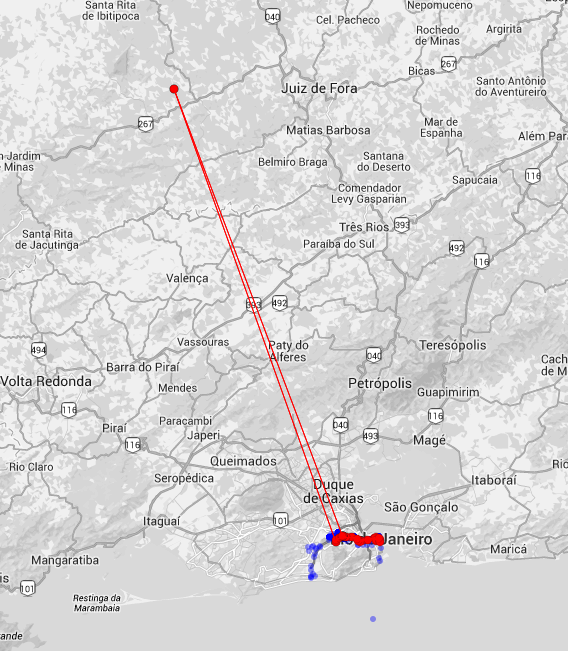}
\caption{\label{fig:gpsnoise}Noise GPS entries detected by the CNN. The buses were following their normal routes but suddenly appeared many miles away from previous points and then came back again to their normal trajectories.}
\end{figure}

When using RioBusData to explore spatial outliers, users can often run into outliers that are likely to be GPS noise. The examples in Figure~\ref{fig:gpsnoise} show two cases of buses 
that were following their normal routes but suddenly appeared many miles away from the bounds of Rio de Janeiro and then back to their regular routes. 

\section{Conclusions and Future Work}
\label{label:conclusions_and_future_work}

Recently, large volumes of data from transportation systems, such as GPS coordinates of public buses, have become available. To understand these data, it is crucial to identify and explore interesting features that arise in the data. 
In this paper,  we presented RioBusData, a visual analytics tool that users  CNN to automatically
detect outliers and provides interactive visual representations which helps users understand and analyze the detected outliers. 

By using a CNN, the task of specifying and detecting outliers is streamlined. While this is more efficient than having users identify the events, the entire process is more opaque. As future 
work, we want to investigate how to mitigate the lack of transparency in the outlier detection by injecting more user feedback into the CNN model.

\addtolength{\textheight}{-12cm}   % This command serves to balance the column lengths

%%%%%%%%%%%%%%%%%%%%%%%%%%%%%%%%%%%%%%%%%%%%%%%%%%%%%%%%%%%%%%%%%%%%%%%%%%%%%%%%

%%%%%%%%%%%%%%%%%%%%%%%%%%%%%%%%%%%%%%%%%%%%%%%%%%%%%%%%%%%%%%%%%%%%%%%%%%%%%%%%

%%%%%%%%%%%%%%%%%%%%%%%%%%%%%%%%%%%%%%%%%%%%%%%%%%%%%%%%%%%%%%%%%%%%%%%%%%%%%%%%

\noindent \textbf{Acknowledgments.}
We thank Luciano Barbosa for providing the bus data set used in this paper. This work was supported in part by a Google Faculty Award, an IBM Faculty Award, the Moore-Sloan Data Science Environment at NYU, DARPA, and NSF awards CNS-1229185, CNS-1405927, and CCF-1533564.

\bibliographystyle{abbrv}
%\addbibresource{busrio.bib}
\bibliography{main}

\end{document}